\documentclass[preprint,12pt]{elsarticle}

\usepackage{amssymb}
\usepackage{url}

\journal{Computational Statistics \& Data Analysis}

\begin{document}

\begin{frontmatter}



\title{Segmentation procedure based on Fisher's exact test and its application to foreign exchange rates}


\author{Aki-Hiro Sato$^{1}$, Hideki Takayasu$^{2}$}

\address{$^{1}$Department of Applied Mathematics and Physics, \\
Graduate School of Informatics, Kyoto University \\
Yoshida Honmachi, Sakyo-ku, 606-8501, Kyoto Japan \\
$^{2}$Sony Computer Science Laboratories, \\
Takanawa Muse Bldg., 3-14-13, Higashigotanda, \\
Shinagawa-ku, 141-0022, Tokyo Japan }

\begin{abstract}
This study proposes the segmentation procedure of univariate time
series based on Fisher's exact test. We show that an adequate change
point can be detected as the minimum value of p-value. It is shown
that the proposed procedure can detect change points for an artificial
time series. We apply the proposed method to find segments of the
foreign exchange rates recursively. It is also applied to randomly
shuffled time series. It concludes that the randomly shuffled data can
be used as a level to determine the null hypothesis.
\end{abstract}


\end{frontmatter}


\section{Introduction}
Recently researchers on data analysis have paid significant attention to
change--points detection. This is a problem to find an adequate
change--points from nonstationary time series. One faces 
this problem when treating data on socio-economic-environmental systems.

The literature on change--point detection is rather huge: reference
monographs include Basseville and Nikiforov~\cite{Basseville}, Brodsky and
Darkhovsky~\cite{Brodsky}, Cs\"org\H{o} and Horv\'ath~\cite{Csorgo},
Chen and Gupta~\cite{Chen:00}. The journal articles by Giraitis and
Leipus~\cite{Giraitis:92,Giraitis:90},
Hawkins~\cite{Hawkins:77,Hawkins:01}, Chen and Gupta~\cite{Chen:04}, 
Mia and Zhao~\cite{Mia}, Sen and Srivastava~\cite{Sen} among others,
are also of interest. 

There are two types of approaches to segmentation procedure. One is a
local approach and another is a global approach. Lavielle and
Teyssi\`ere~\cite{Lavielle} addressed the issue of global procedure vs
local procedure, and found that the extension of single change--points
procedures to the case of multiple change--point using
Vostrikova\'s~\cite{Vostrikova} binary segmentation procedure is
misleading and yields an overestimation of the number of
change--points. Quantdt and Ramsey have developed estimation procedure
for mixture of linear regression~\cite{Quandt}. Kawahara and
Sugiyama~\cite{Kawahara} propose a non-parametric method to detect
change points from time series based on direct density--ratio
estimation. More recently, a recursive entropic scheme to separate
financial time series has been proposed~\cite{Cheong}. Their method is
parametric and uses the log--likelihood ratio test.

Decr\'e-Robitaille et al. ~\cite{Ducre} compare several methods to
detect change points. They segment artificial time series based on 8
methods; standard normal homogeneity test (SNHT) without
trend~\cite{Alexandersson:86}, SNHT with
trend~\cite{Alexandersson:97}, multiple linear regression
(MLR)~\cite{Vincent}, two-phase regression (TPR)~\cite{Easterling},
Wilcoxon rank-sum (WRS)~\cite{Karl}, sequential testing 
for equality of means (ST)~\cite{Gullett}, Bayesian approach without
reference series~\cite{Perreault:99,Perreault:00}, and Bayesian approach with
reference series~\cite{Perreault:99,Perreault:00}. Karl and
Williams~\cite{Karl} propose a method to find an adequate segment boundary
based on Wilcoxon rank--sum test and investigate climatological time
series data.

We further address some existing approaches to the problem of multiple
change--point detection in multivariate time series. Ombao et
al.~\cite{Ombao:05} employ the SLEX (smooth localized complex
exponentials) basis for time series segmentation, originally proposed
by Ombao et al.~\cite{Ombao:02}. The choice of SLEX basis leads to the 
segmentation of the time series, achieved via complexity--penalized
optimization. Lavielle and Teyssi\`ere~\cite{Levielle:06} introduce a
procedure based on penalized Gaussian log--likelihood as a cost
function, where the estimator is computed via dynamic
programming. The performance of the method is tested on bivariate
examples. Vert and Bleakley~\cite{Vert} propose a method for
approximating multiple signals (with independent noise) via piecewise
constant functions, where the change-point detection problem is
re-formulated as a penalized regression problem and solved by the
group Lasso~\cite{Yuan}. Note that Cho and Fryzlewicz~\cite{Cho}
argue that Lasso-type penalties are sub-optimal for change-point
detection.

In this study, we propose a non-parametric method to detect an
adequate segment boundary. Our procedure is based on Fisher's exact test
and uses it as a discriminant measure to detect the segmentation
boundary. Fisher's exact test can be used in the case of a small
number of samples. The proposed method is performed with both artificial
nonstationary time series and actual data on daily log--return time
series.

This paper is organized as follows. We introduce Fisher's exact test and
propose a non-parametric method to find an adequate change point from
time series in Section \ref{sec:Fisher}. We perform the proposed method
with artificial nonstationary time series in Section \ref{sec:numerical-study}.
We conduct empirical analysis of the daily log--return time series of AUD/JPY
in Section \ref{sec:empirical-analysis}. Section \ref{sec:conclusion}
is devoted to conclusions.

\section{Fisher exact test for segmentation procedure}
\label{sec:Fisher}
We want to find an adequate segment regarding
non-stationarity of $r(t)$. Assume that the time series $r(t)$ consists of
$n$ segments. The problem is to determine $n-1$ boundaries from the time 
series.

Firstly, we discuss a method to determine an adequate boundary in a time series
based on Fisher's exact test. Introducing $x_{th}$ in a value and $\tau$
in time, we count the number of four events:
\begin{itemize}
\item $a$ : the number of $r(t) > x_{th}$ for $0 \leq t < \tau$.
\item $b$ : the number of $r(t) \leq x_{th}$ for $0 \leq t < \tau$.
\item $c$ : the number of $r(t) > x_{th}$ for $\tau < t \leq T$.
\item $d$ : the number of $r(t) \leq x_{th}$ for $\tau < t \leq T$.
\end{itemize}
According to Fisher's exact test, fixing the vertical numbers $a+c$ and
$b+d$, horizontal numbers $a+b$ and $c+d$, the two-sided probability
of $2 \times 2$ matrix $\left[\begin{array}{cc}a & b \\ c &
    d \end{array}\right]$ is given as
\begin{equation}
p(a,b,c,d) = \frac{\sum_{i=0}^{\mbox{min}(b,c)}(_{a+c} C_{c-i} \cdot _{b+d} C_{b-i})
  + \sum_{i=0}^{\mbox{min}(a,d)} (_{a+c} C_{a-i} \cdot _{b+d}
  C_{d-i})}{_n C_{a+b}}
\label{eq:pabcd}
\end{equation}
This probability is used as p-value for realized quartet $(a,b,c,d)$. 
The smallest p-value tells us that the threshold $x_{th}$ and the
boundary $\tau$ are of statistically significance. Therefore, we can
detect a change point from the minimization problem; 
\begin{equation}
\{\hat{\tau}, \hat{x}_{th}\} = \arg \min_{\tau, x_{th}} p\Bigl(a(\tau,x_{th}),b(\tau,x_{th}),c(\tau,x_{th}),d(\tau,x_{th})\Bigr).
\end{equation}

If the minimum p-value for the event which we observed as the quartet
$(a,b,c,d)$ is less than a certain significance level $p_{th}$, then
the event is regarded as a statistically significant case and the
boundary $\tau$ should be accepted as a change--point.

Let us now briefly discuss some issues related to the determination of
$\hat{\tau}$. The interpretation of this time point $\hat{\tau}$ is
that it gives an optimal separation of the time series into two
statistically most distinct segments. The segmentation procedure can
be used recursively to separate further the segmented time series into
smaller segments. We do this iteratively until the iteration is
terminated by a stopping condition. Several termination conditions
have been proposed in previous studies. We terminate the iteration if
the p-value is larger than the amplitudes of typical fluctuations in
the spectra. 

In practice,  in order to simplify the recursive segmentation, we
adopt a conservative threshold of $p_{th}$ and terminate the procedure
if $p(a,b,c,d)$ is less than $p_{th}$. 

\section{Numerical Study}
\label{sec:numerical-study}
\begin{figure}[hbt]
\centering
\includegraphics[scale=0.8]{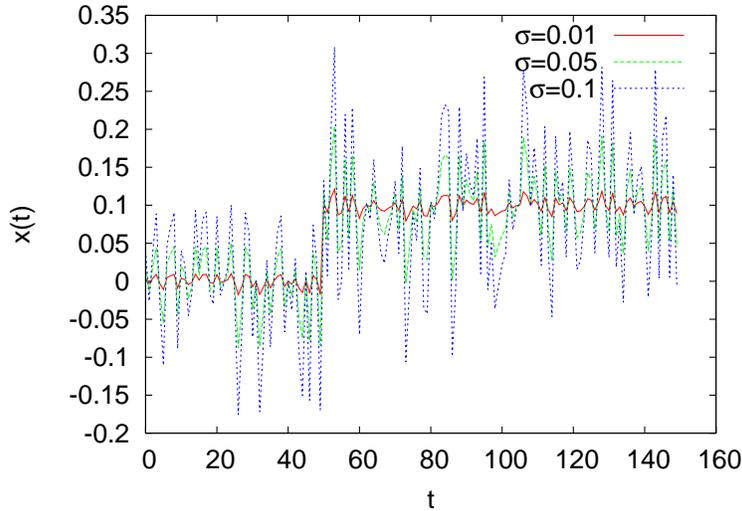}
\caption{Artificial nonstationary time series at $\sigma=0.01$, $0.05$, and $0.1$ with the same random seed.}
\label{fig:numerical}
\end{figure}

As artificial nonstationary time series, we generate time series as the 
following algorithm.
\begin{equation}
x(t) = 
\left\{
\begin{array}{ll}
\sigma \xi(t) & 1 \leq t \leq 50 \\
\sigma \xi(t)+0.1 & 51 \leq t \leq 150 \\
\end{array}
\right.,
\end{equation}
where $\xi(t)$ is drawn from {\it i.i.d.} standard normal distribution with
zero mean, and standard deviation $\sigma$. Fig. \ref{fig:numerical}
shows a sample of artificial nonstationary time series with
$\sigma=0.01$, $0.05$, and $0.1$. These time series are generated from
the same random seed.

\begin{figure}[hbt]
\centering
\includegraphics[scale=0.6]{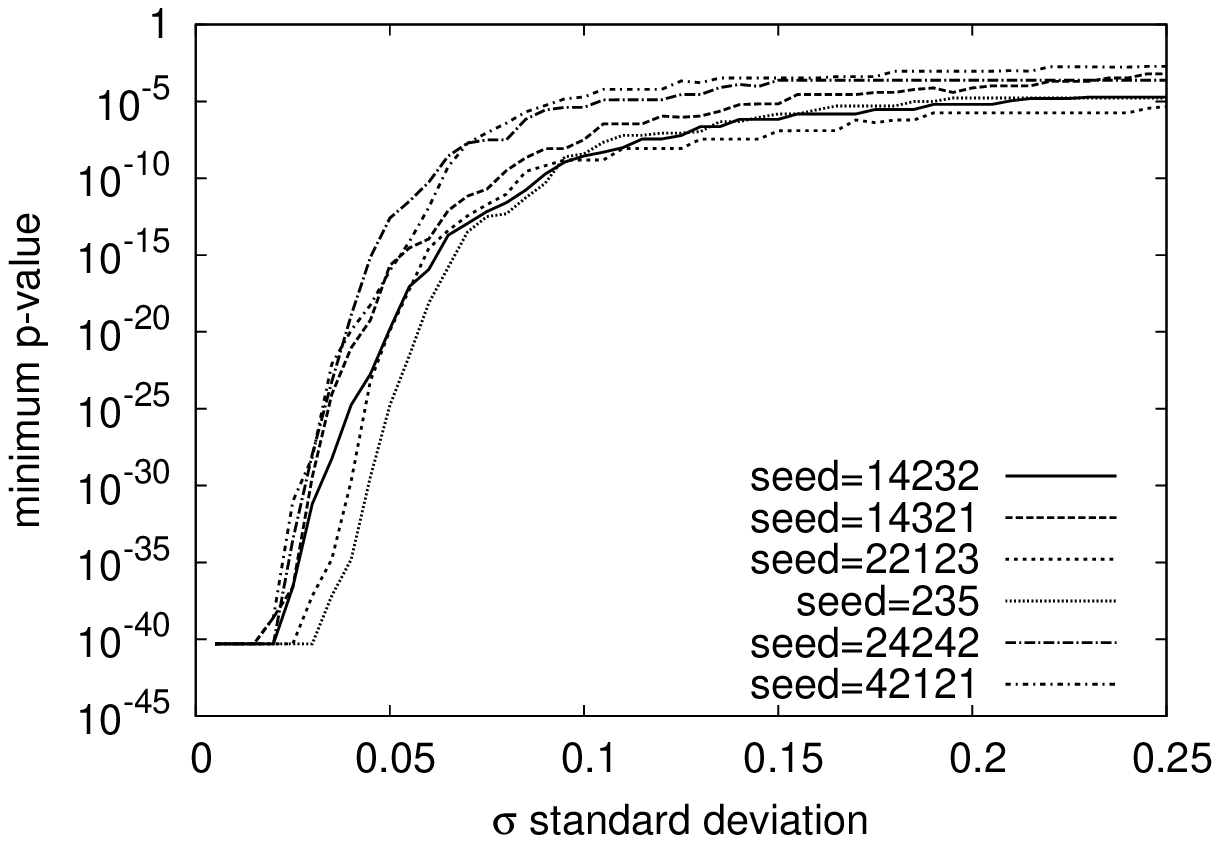}(a)
\includegraphics[scale=0.6]{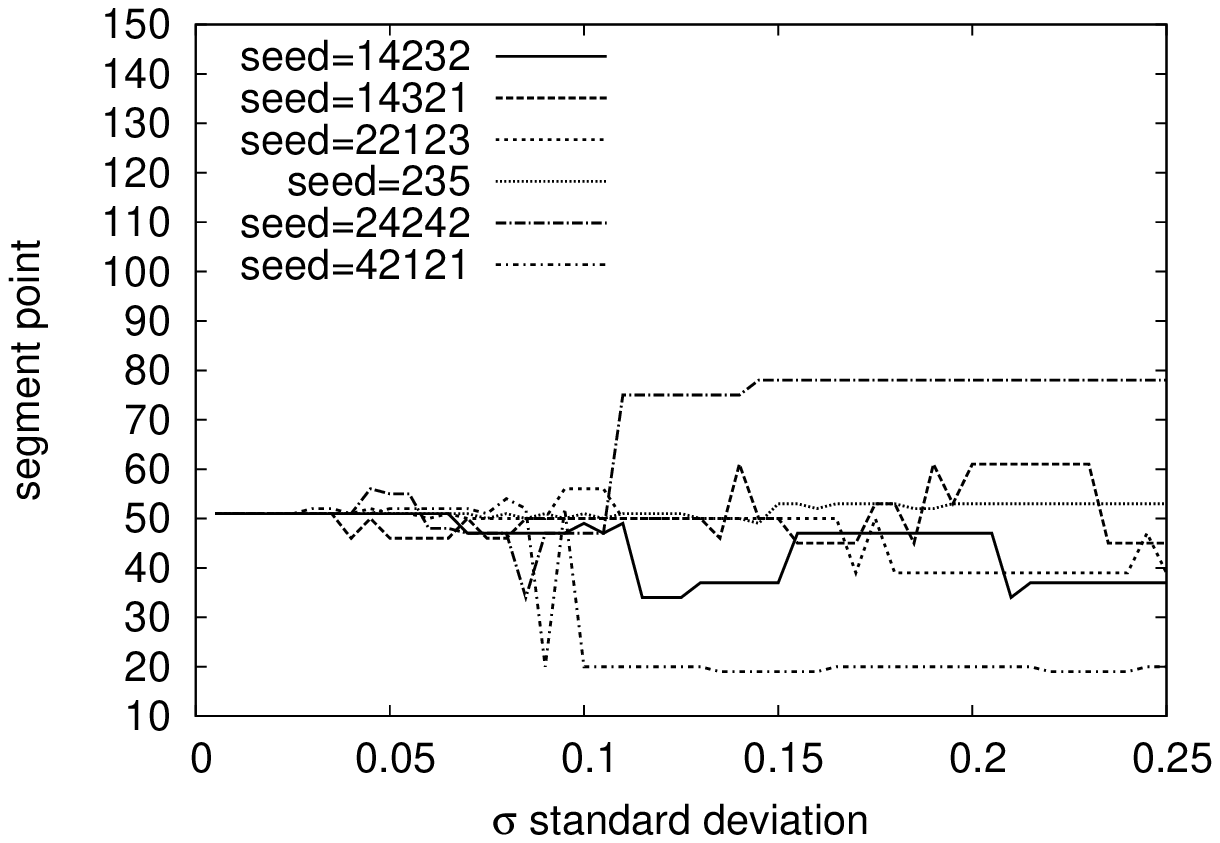}(b)
\caption{(a) The relationship between $\sigma$ and the minimum p-value. Each
curve corresponds to the relationship between $\sigma$ and the minimum p-value
computed from the artificial time series with the same random seed.
(b) The points giving the minimum p-value at $\sigma$. Each curve corresponds
to the relationship between $\sigma$ and the estimated change point.}
\label{fig:error}
\end{figure}

We compute the minimum p-value varying the value of $\sigma$. 
Fig. \ref{fig:error} shows the relationship between the minimum
p-value and $\sigma$. As $\sigma$ increases, the minimum p-value
increases. Namely, the estimation error for change points increases as
$\sigma$ increases. In this case, statistically significant estimation
of the change point is given for $\sigma < 0.1$. This implies that a
point where the minimum p-value is less than $10^{-5}$ is
statistically significant.

\begin{figure}[hbt]
\centering
\includegraphics[scale=0.8]{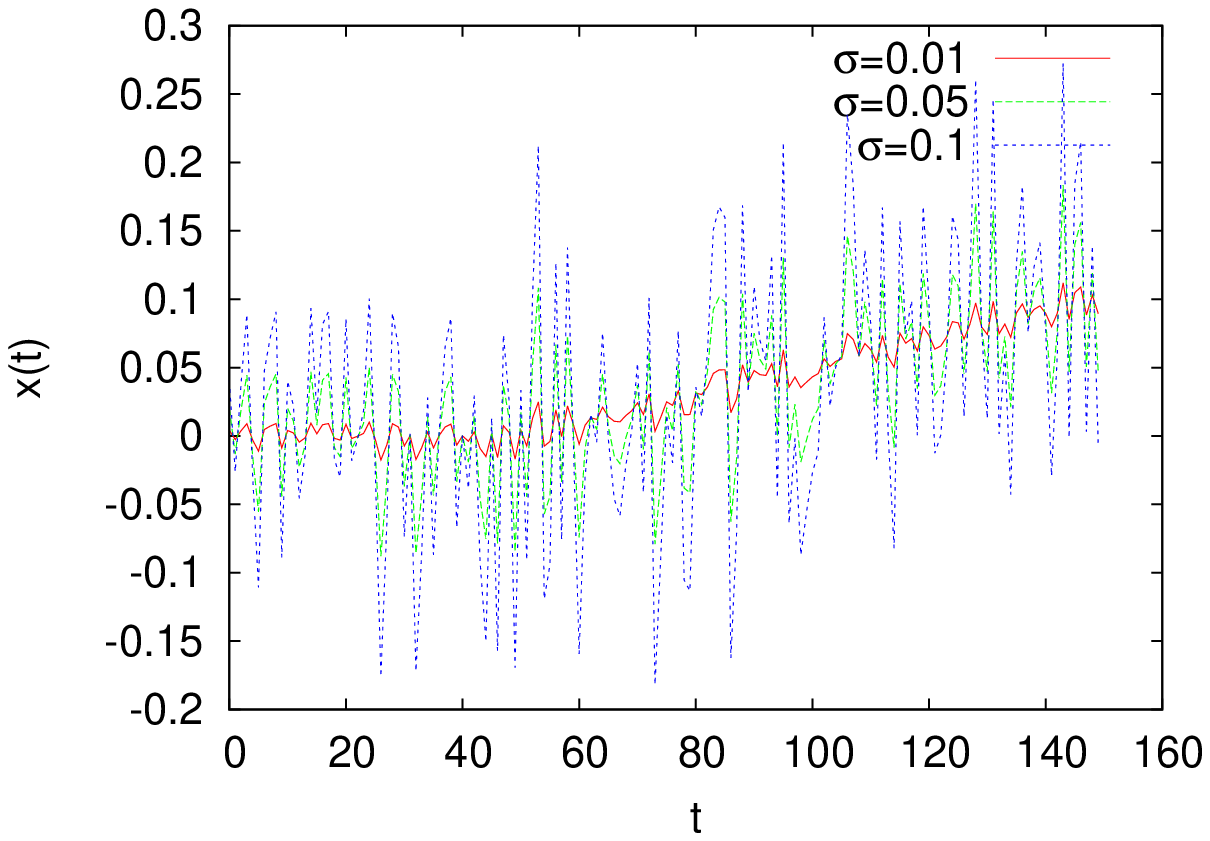}
\caption{Artificial nonstationary time series at $\sigma=0.01$, $0.05$, and $0.1$ with the same random seed.}
\label{fig:numerical-2}
\end{figure}

As artificial nonstationary time series, we generate another type of time
series as the following algorithm.
\begin{equation}
x(t) =
\left\{
\begin{array}{ll}
\sigma \xi(t) & 1 \leq t \leq 50 \\
\sigma \xi(t)+0.001(t-50) & 51 \leq t \leq 150 \\
\end{array}
\right.,
\end{equation}
where $\xi(t)$ is drawn from {\it i.i.d.} standard normal distribution with
zero mean, and standard deviation $\sigma$. Fig. \ref{fig:numerical-2}
shows artificial nonstationary time series with $\sigma=0.01$, $0.05$,
and $0.1$. These time series are generated from the same random seed. 

\begin{figure}[hbt]
\centering
\includegraphics[scale=0.6]{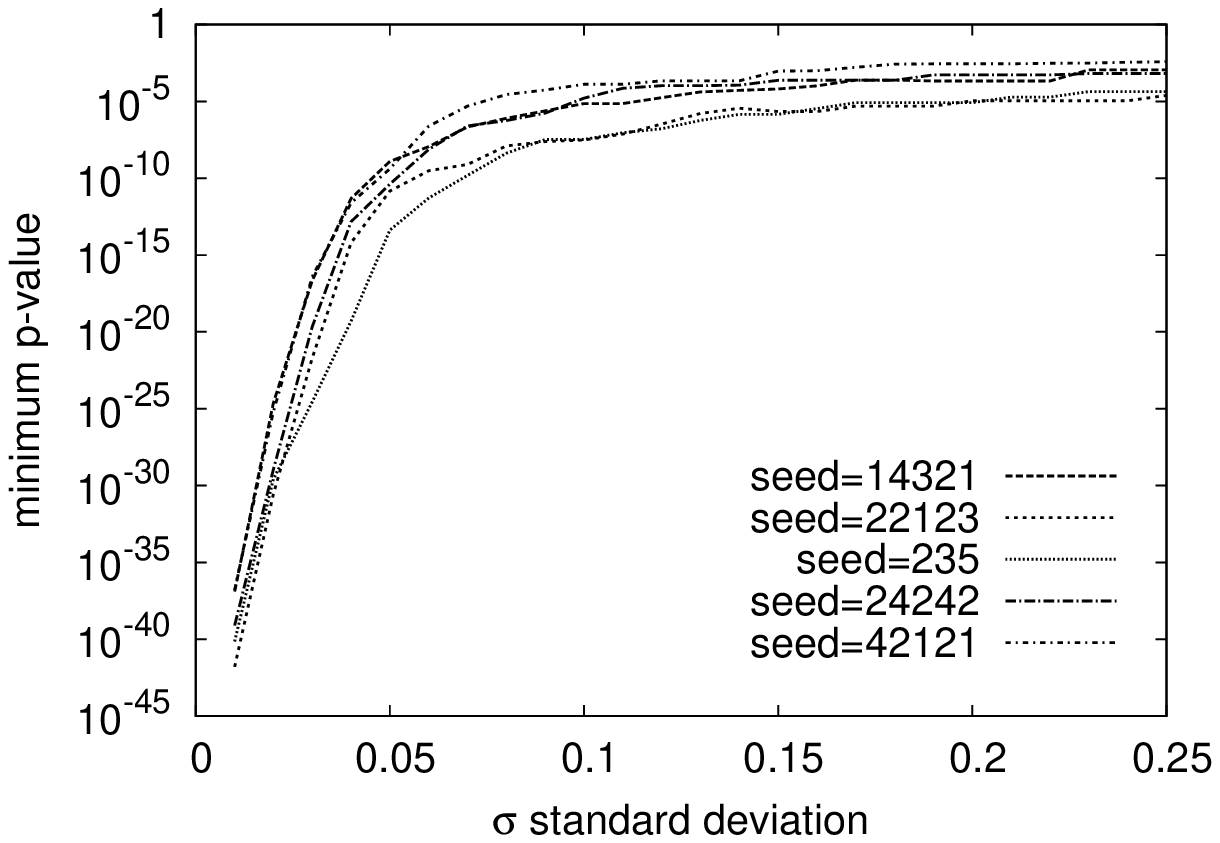}(a)
\includegraphics[scale=0.6]{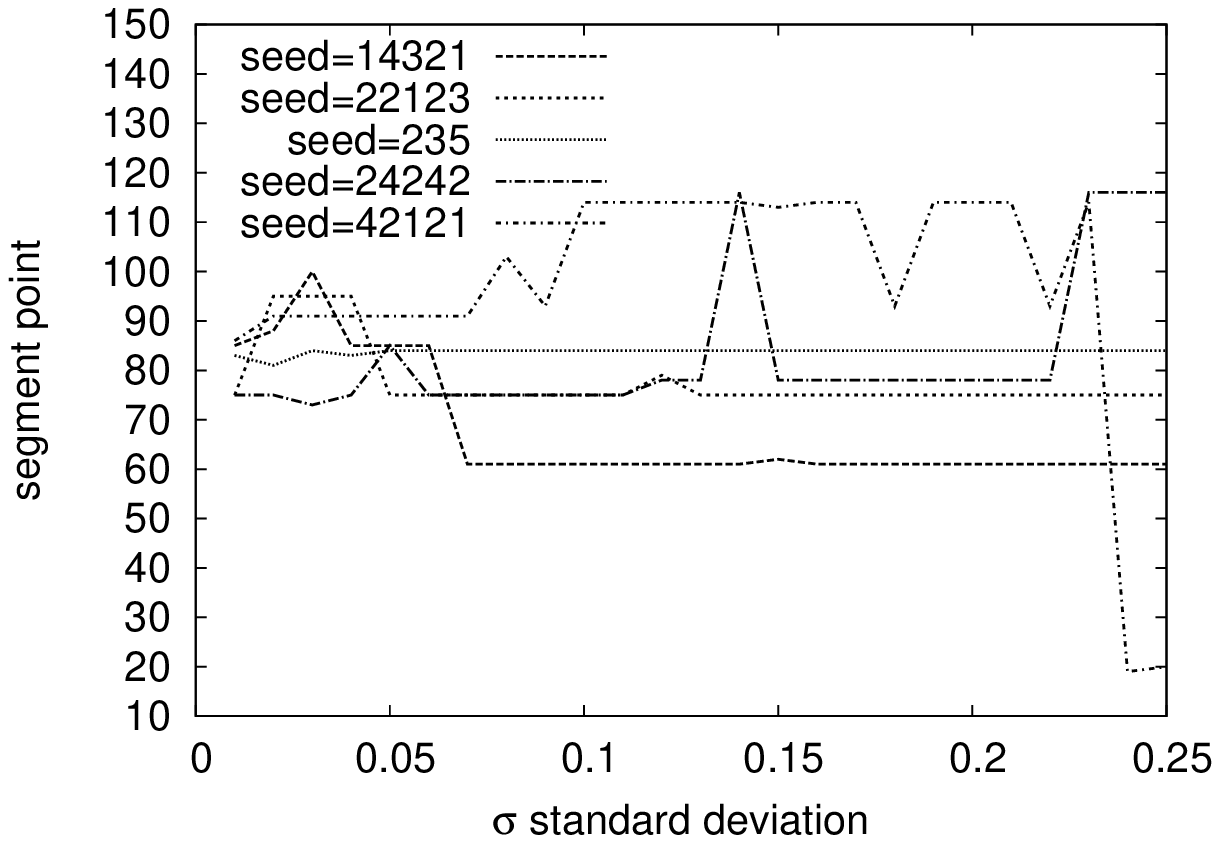}(b)
\caption{(a) The relationship between $\sigma$ and the minimum p-value. Each
curve corresponds to the relationship between $\sigma$ and the minimum p-value
computed from the artificial time series with the same random seed.
(b) The points giving the minimum p-value at $\sigma$. Each curve corresponds
to the relationship between $\sigma$ and the estimated change point.}
\label{fig:error-2}
\end{figure}

We compute the minimum p-value varying the value of $\sigma$. 
Fig. \ref{fig:error-2} shows the relationship between the minimum p-value and $\sigma$. As $\sigma$ increases, the minimum p-value increases. Namely,
the estimation error for change points increases as $\sigma$ increases. In 
this case, statistically significant estimation of the change point is given
for $\sigma < 0.2$. This implies that a point where the minimum p-value 
is less than $10^{-5}$ is statistically significant. 

Comparing p-values from two examples, p-values shown in
Fig. \ref{fig:error} show smaller values than those in
Fig. \ref{fig:error-2}. The error sensitivity depends on the type of
nonstationarity.

\section{Empirical Analysis}
\label{sec:empirical-analysis}
We apply a recursive segmentation procedure with our proposed method
to an actual time series. We use daily log--return time series of
foreign exchange rate of Australian Dollar (AUD) against Japanese Yen
(JPY). The log--return time series of AUD/JPY for the period from 03
January, 2001 to 30 December, 2011. Throughout this analysis, we set
the threshold as  $p_{th}=10^{-5}$.

Suppose that $R(t)$ represents an exchange rate at business day 
$t \quad (t=1,\ldots,n+1)$. Let $r(t)$ be a daily log--return of exchange rate 
at time $t \quad (t=1,\ldots,n)$, which is defined as $r(t) = \log
R(t+1) - \log R(t)$. Following the procedure shown in
Sec. \ref{sec:Fisher}, we count the number of four events and compute
the p-value. 

\begin{figure}[!phbt]
\centering
\includegraphics[scale=0.7]{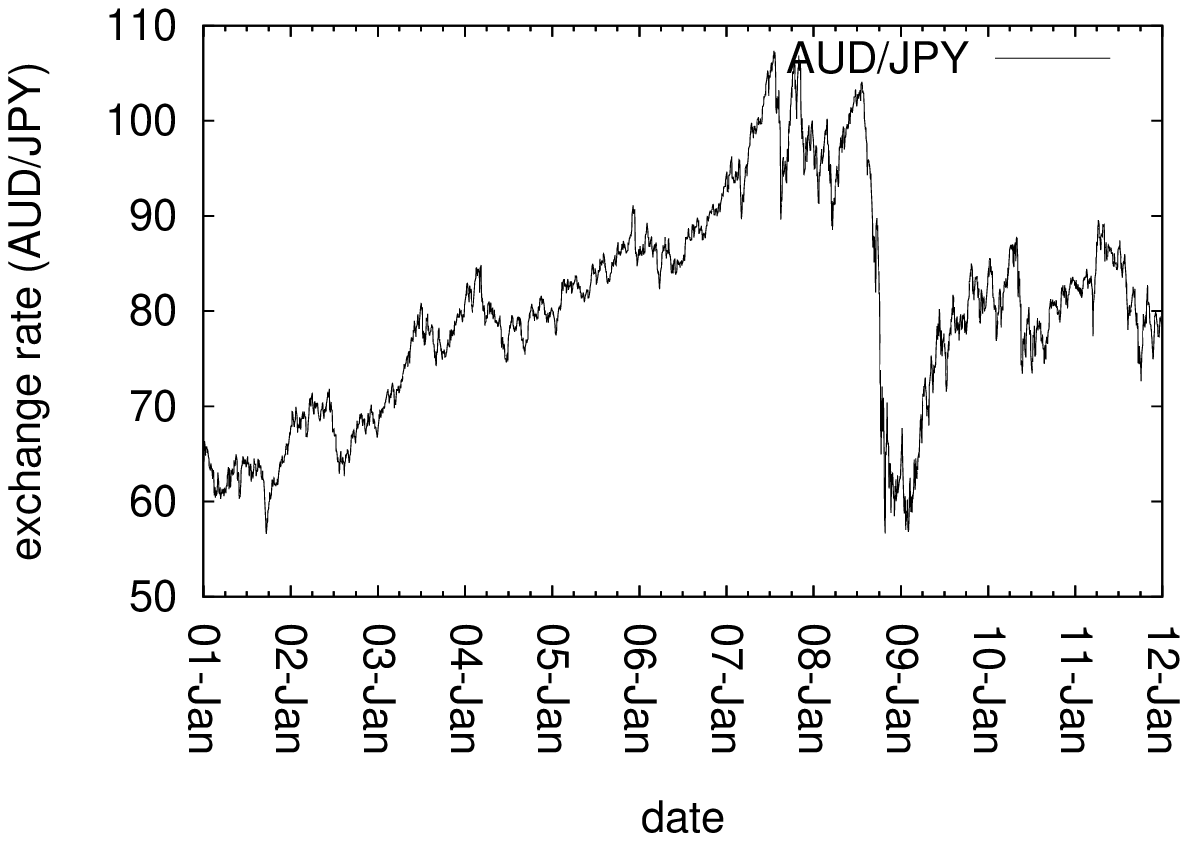}(a)
\includegraphics[scale=0.7]{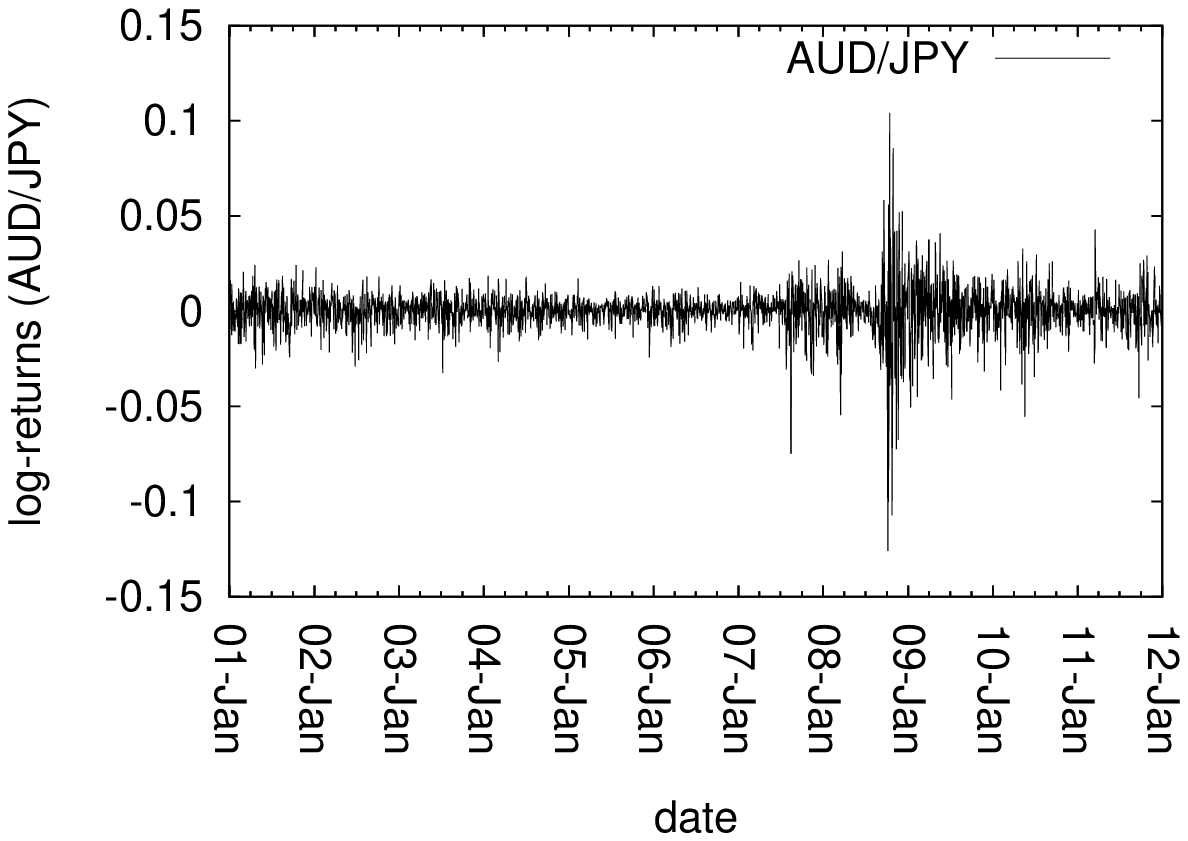}(b)
\includegraphics[scale=0.7]{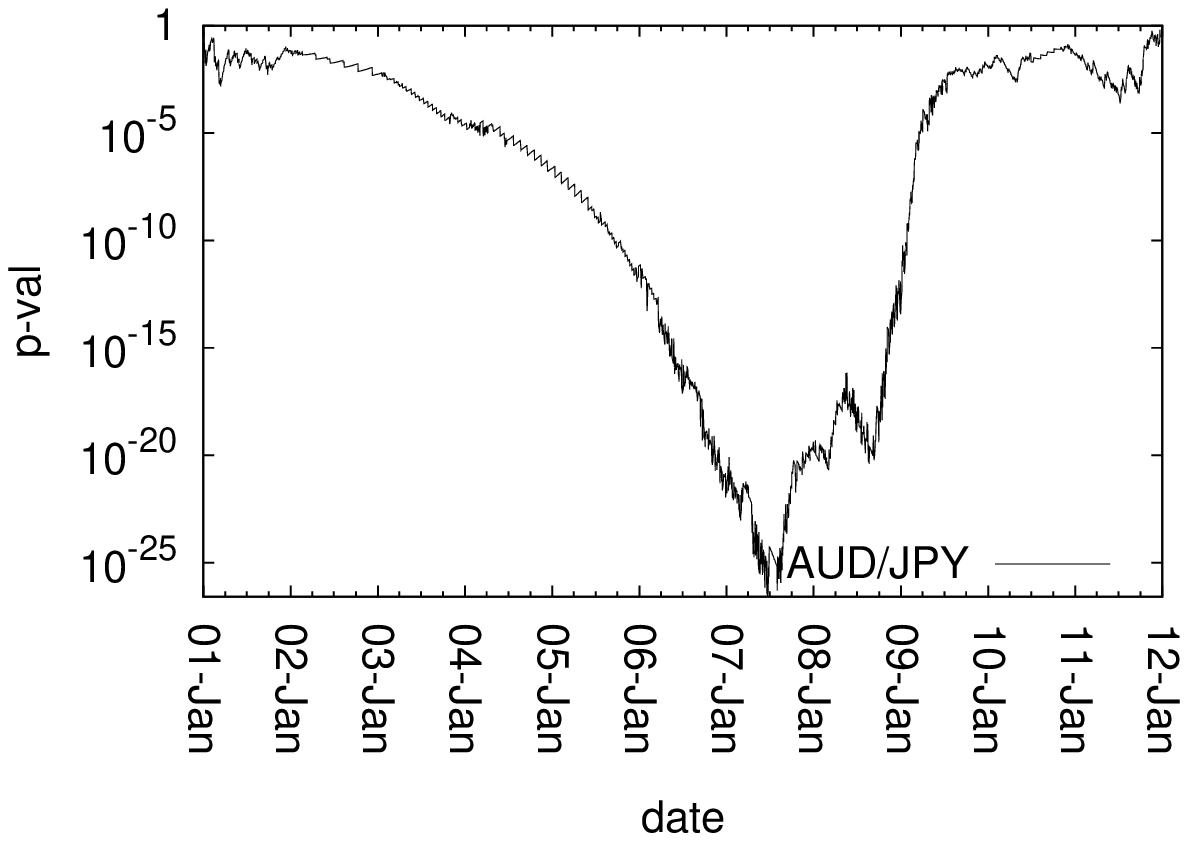}(c)
\caption{(a) Daily exchange rates of AUD/JPY, (b) daily log return
  time series of AUD/JPY, (c) the p-value computed from the daily
  log--return time series of AUD/JPY with Fisher exact.} 
\label{fig:AUDJPY}
\end{figure}

Fig. \ref{fig:AUDJPY} shows exchange rates of AUD/JPY, their daily
log--return time series, and the p-value computed from Fisher' exact
test. We compute the minimum p-value changing $x_{th}$ from
Eq. (\ref{eq:pabcd}) on business day $\tau$. From Fig. \ref{fig:AUDJPY}
(c) we observe the smallest p-value, which is estimated as 
$2.6\times 10^{-27}$, on 22 June, 2007. Since the smallest p-value is less
than $p_{th}=10^{-5}$, we determine this date as an adequate
boundary. Repeating to apply this procedure to each two segments, we
obtain all the segments until the smallest p-value is greater than $p_{th}$.
We detect six segments in the time series and show sample mean and
sample standard deviation computed from each segment in
Tab. \ref{tab:segments}.

Suppose that we have $K$ segments. Let $[t_{k-1}+1, \ldots, t_k]$
represent a range of the $k$-th segment ($t_0=0$). The mean value of
the $k$-th segment is computed as
\begin{equation}
\mu^{(k)} = \frac{1}{t_{k}-t_{k-1}}\sum_{t=t_{k-1}+1}^{t_k} r(t).
\end{equation}
We attempt to obtain an adequate fitting curve in each segment 
based on each estimated segment. Since we can describe
$R(t+1) = R(t) \exp\bigl(r(t)\bigr)$ of exchange rate $R(t)$ at
business time $t$, we assume that the exchange rate is fitted with 
\begin{equation}
R(t) = \exp\Bigl(\mu^{(k)} (t-t_{k-1}) + \rho \Bigr) \quad (t_{k-1}+1
\leq t \leq t_{k}),
\end{equation}
The value $\rho$ is estimated with the least squared method:
\begin{equation}
\rho = \frac{1}{t_k-t_{k-1}}\sum_{t=t_{k-1}+1}^{t_k} \Bigl( \log R(t)
- \mu(t-t_{k-1}) \Bigr) 
\label{eq:bb}
\end{equation}
We can compute parameter $\rho$ of this simple regression model from
time series. The detail derivation of Eq. (\ref{eq:bb}) is described
in \ref{sec:bb}.

\begin{figure}
\centering
\includegraphics[scale=0.7]{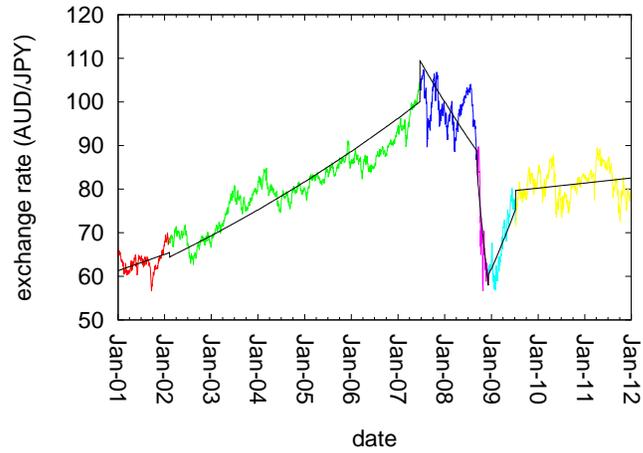}(a)
\includegraphics[scale=0.7]{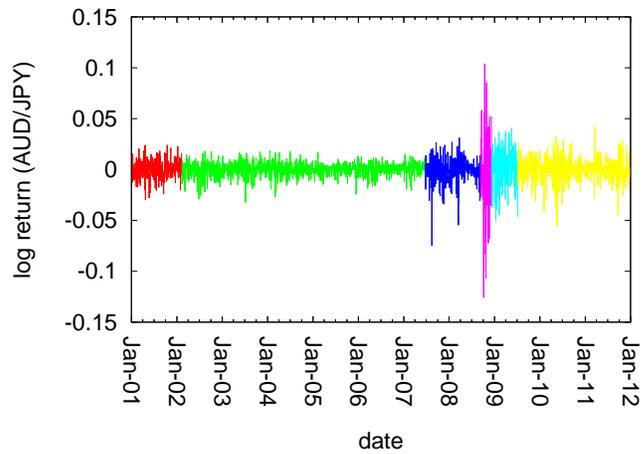}(b)
\caption{(a) Daily exchange rates of AUD/JPY segmented by the recursive
Fisher Exact test and (b) segmented daily log return of AUD/JPY.}
\label{fig:segments}
\end{figure}

A positive trend of AUD against JPY reveals in both the first and
second segments. In the third segment AUD against JPY drops slightly and in the
fourth segment the exchange rates of AUD against JPY steeply drops due to
the influence of global financial crisis driven by the Lehman shock. In both
the fifth and sixth segments the exchange rates of AUD/JPY are eventually 
stabilized and volatility decreases. Fig. \ref{fig:segments} shows the
segments detected with the proposed method. The first and second
segments are related to a growth phase in the world economy. In the
third segment, we observed instability of the global economy. The
fourth segment coincides the global financial crisis triggered by
bankruptcy of Lehman brothers in September 2008. After the Lehman
shock, JPY rapidly dropped against AUD.

\begin{figure}[!phbt]
\centering
\includegraphics[scale=0.7]{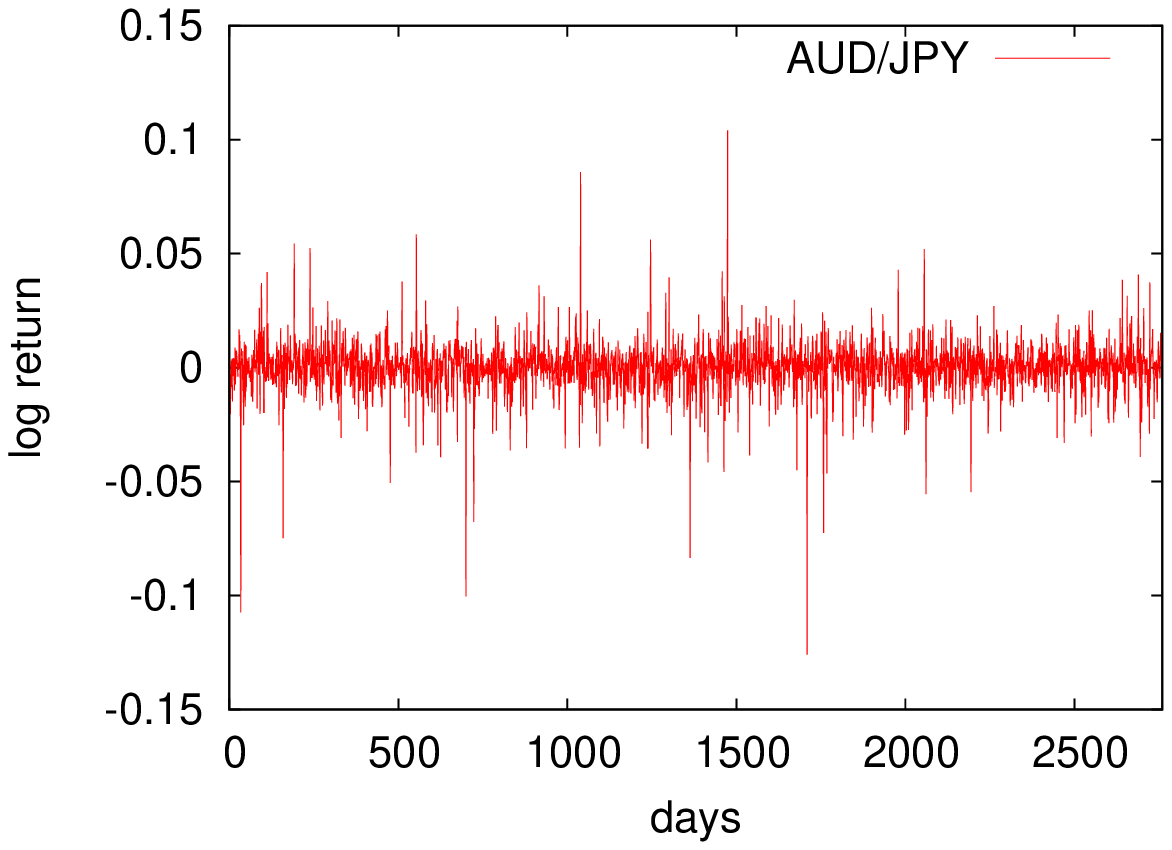}(b)
\includegraphics[scale=0.7]{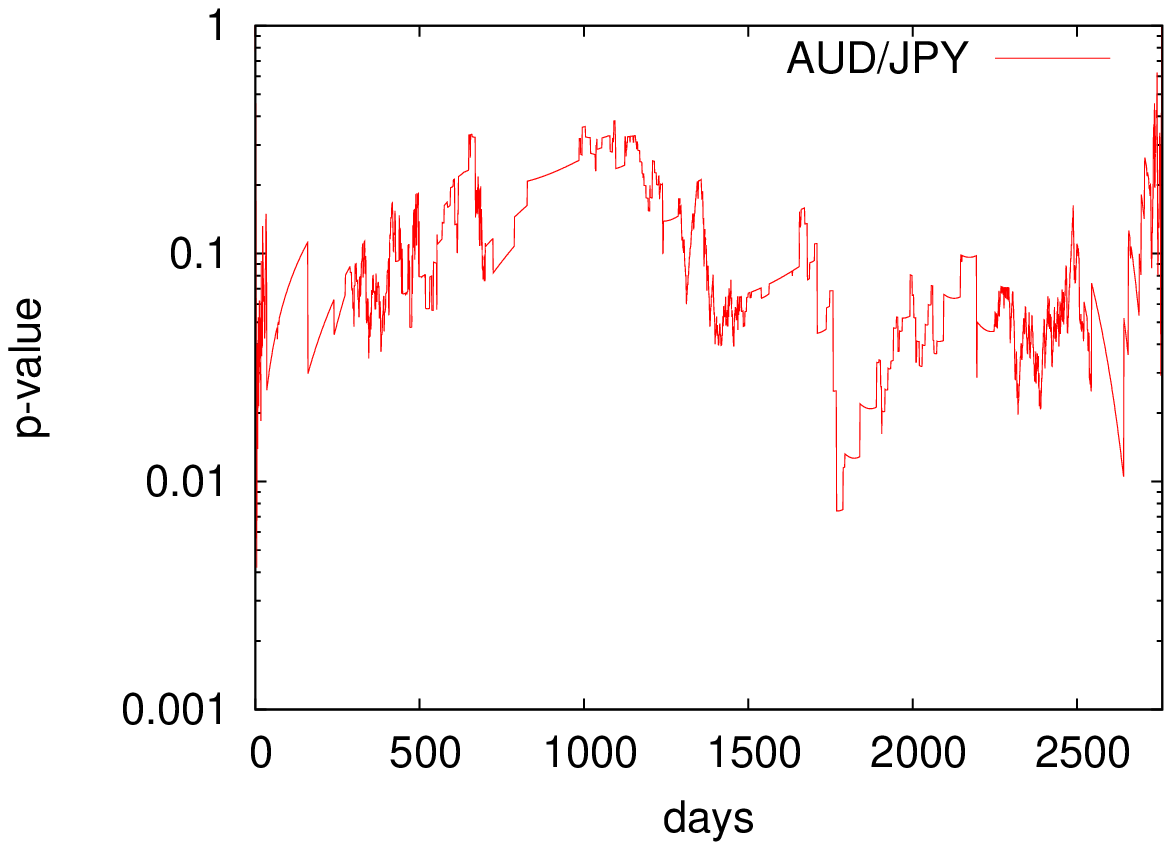}(c)
\caption{(a) Randomized daily log return time series of AUD/JPY, and 
(b) the p-value computed from daily log return time series.}
\label{fig:RS}
\end{figure}

We generate randomly shuffled data from the log--return time
series. The randomly shuffled data preserves sample mean and
sample standard deviation of original time series. We compute p-value
for the randomly shuffled data with the proposed
method. Fig. \ref{fig:RS} (a) shows the randomly shuffled time series,
and (b) p-value computed for the randomly shuffled data. The p-value
is almost always larger than $10^{-2}$. This implies that the p-value
of original time series is more statistically significant than
randomly shuffled data.

\begin{table}[!hbt]
\caption{Descriptive statistics of each segment.}
\label{tab:segments}
\centering
\begin{tabular}{lllll}
\hline
no. & start & end & mean & std. \\
\hline
1 & 2001-1-3 & 2002-2-8 & 0.000253 & 0.009131 \\
2 & 2002-2-11 & 2007-6-21 & 0.000312 & 0.006308 \\
3 & 2007-6-22 & 2008-9-11 & -0.000676 & 0.012243 \\
4 & 2008-9-12 & 2008-12-8 & -0.005385 & 0.042808 \\
5 & 2008-12-9 & 2009-7-8 & 0.001095 & 0.018160 \\
6 & 2009-7-9 & 2011-12-30 & 0.000141 & 0.010740 \\
\hline
\end{tabular}
\end{table}

\section{Conclusion}
\label{sec:conclusion}
This paper proposed the segmentation procedure based on Fisher's
exact test. The proposed method was tested for artificial time
series. It was confirmed that the ratio of mean values of time series
to their volatility is associated with the estimation error of
segmentation boundary. As the variance increases, the estimation error
increases. The proposed method was applied to the actual data of daily
foreign exchange rates. The change points detected with the proposed
method were characterized in terms of sample mean and standard
deviations of daily log--return time series. The randomly shuffled
data was used as the null hypothesis. The shuffled data showed the
large p-value which can not be accepted as a segmentation boundary.

\appendix
\section{Derivation of the price fitting curve}
\label{sec:bb}
Assume that $R(t) \quad (t=1,\ldots,n+1)$ is divided into $K$
segments. The $k$-th segment takes a range from $t_{k-1}+1$ to
$t_{k}$, where $t_0 = 0$ and $t_{K+1}=n+1$. Suppose the fitting curve
of the $k$-th segment of exchange rate $R(t) \quad (t=t_{k-1}+1,\ldots,t_k)$ as
\begin{equation}
R(t) = \exp\Bigl(\mu(t-t_{k-1}) + \rho \Bigr).
\end{equation}
We assume that the least squared error can be written as
\begin{equation}
E(\rho) = \sum_{t=t_{k-1}+1}^{t_k} \Bigl(\log R(t) - \mu(t-t_{k-1}) - \rho \Bigr)^2.
\end{equation}
Differentiating $E(\rho)$ in terms of $\rho$ and setting it into zero,
we get
\begin{equation}
2 \sum_{t=t_{k-1}+1}^{t_k} \Bigl( \log R(t) - \mu(t-t_{k-1}) - \rho \Bigr) = 0.
\end{equation}
Consequently, we obtain
\begin{equation}
\nonumber
\rho = \frac{1}{t_k-t_{k-1}} \sum_{t=t_{k-1}+1}^{t_k} \Bigl( \log R(t) -
\mu (t - t_{k-1}) \Bigr).
\end{equation}








\end{document}